\documentstyle[12pt]{article}
\def\gapp{\mathrel{\raise.3ex\hbox{$>$}\mkern-14mu
              \lower0.6ex\hbox{$\sim$}}}
\def\gsim{\gapp}
\def\lapp{\mathrel{\raise.3ex\hbox{$<$}\mkern-14mu
              \lower0.6ex\hbox{$\sim$}}}
\begin{document}
\begin{titlepage}
\begin{center}
\today     \hfill    MIT-CTP-2350 \\
           \hfill    hep-ph/9409419 \\
\vskip .5in

{\large \bf Measures of Fine Tuning}
\footnote{This work was supported in part by funds provided by the
U.S. Department of Energy (DOE) under cooperative agreement
DE-FC02-94ER40818 and
by the Texas National Research Laboratory Commission under grant
RGFY932786.}
\vskip .5in
Greg W. Anderson \footnote {email address: anderson@fnth03.fnal.gov}
and Diego J. Casta\~no \footnote
{email address: castano@fshewj.hep.fsu.edu}\\
{\em Center for Theoretical Physics \\
     Laboratory for Nuclear Science \\
     Massachusetts Inst. of Technology\\
     Cambridge, MA 02139.}\\

\end{center}

\vspace{\fill}
\begin{center}
{\it To appear in:} Physics Letters {\bf B347}, 300-308 (1995)
\end{center}
\vskip .5in

\begin{abstract}
Fine-tuning criteria are frequently used to place upper limits
on the masses of superpartners in supersymmetric extensions
of the standard model.  However, commonly used prescriptions for
quantifying naturalness have some important shortcomings.
Motivated by this, we propose new criteria for
quantifying fine tuning that can be used to place upper limits on
superpartner masses with greater fidelity.  In addition, our analysis
attempts to make explicit the assumptions implicit in
quantifications of naturalness.
We apply our criteria to the minimal supersymmetric extension
of the standard
model, and we find that the scale of supersymmetry breaking can
be larger than previous methods indicate.
\end{abstract}
\end{titlepage}

\renewcommand{\thepage}{\arabic{page}}
\setcounter{page}{1}

\section{Introduction}
\def\theequation{1.\arabic{equation}}
\setcounter{equation}{0}
\indent

   One of the principle motivations for weak scale supersymmetry
is that it provides a framework that stabilizes the hierarchy
between the weak scale and the Planck scale, or some other
unification scale.
In  non-supersymmetric models, the mass renormalization of
fundamental scalars is quadratically divergent.
This divergence must be cancelled, 
or the fundamental scalar will have a
renormalized mass on the order of the cutoff.  In the standard model,
if the Higgs boson remains a fundamental degree of freedom all the way
up to some very heavy scale, we must fine tune a precise cancellation
order by order in perturbation theory to maintain the lightness of
the weak scale.

Supersymmetry solves this problem because the renormalization effects
of superpartners eliminate the quadratic divergences.   But
supersymmetry is at best a broken symmetry.
There are no superpartners degenerate in
mass with the particles that have been observed so far.
These superpartners can have gauge invariant mass terms
if supersymmetry is softly broken, and these masses can be made
arbitrarily large provided we increase the scale of supersymmetry
breaking.  There is a price for this.
As the scale of supersymmetry breaking  increases
the weak scale can only remain light by virtue of an increasingly delicate
cancellation.  Eventually a point is reached when the model no longer
appears to provide a complete explanation of why
a light weak scale is stable.

Attempts to pinpoint where and when our understanding
of weak scale stability is lost, or becomes
incomplete, must of necessity quantify some intuitive
notion of naturalness.  Such a prescription for quantifying naturalness
exists and is widely used in the literature.
If we demand that supersymmetric extensions of the standard model
should be ``complete'' in their explanations of this stability,
we can place an upper limit on the scale of
supersymmetry breaking.  This can be translated into
an upper limit on the masses of superpartners.

In this paper, we examine
the prescription that is currently used to place upper bounds
on superpartner masses.
\footnote{Heavy superpartner masses can also be bounded, or
at least restricted, by the requirement that the relic density
of LSP's does not over close the universe.  These constraints
provide interesting limits, but they don't provide an absolute upper
limit on sparticle masses, and they involve model dependent
assumptions concerning conserved R-parity.}
First, we wish to determine if these criteria accurately measure
fine tuning.  Second, we want to make explicit the
assumptions implicit in any attempt to quantify naturalness.
Upper limits on sparticle masses obtained from naturalness criteria
influence 
expectations of when and where sparticles will be discovered
if supersymmetry is responsible for the stability of the weak scale.

  In section two we make a critical examination of fine tuning, and
we analyze the prescription now used to quantify naturalness.
We critique this traditional method by examining a well known
hierarchy.  We find that this prescription is not completely
satisfactory.  The trouble is that the traditional prescription
does not distinguish between instances
of global sensitivity and real instances of fine tuning.
We argue that a reliable measure of fine tuning
requires global information about the dependence of
certain quantities on their arguments, and we show how the existing
prescription
can be augmented with this information to yield reliable measures
of fine tuning.

In section three we systematically construct a family of prescriptions
that coincide with the augmented prescriptions formulated in
section two.  Our construction clarifies the proper normalization
of naturalness measures and
makes explicit the extent of theoretical prejudice present in
any such measure.

In section four we apply our prescription to the minimal supersymmetric
standard model (MSSM).  We briefly discuss the level of fine
tuning the MSSM requires in light of current experimental constraints,
and we show how the current situation is much less
fine tuned than it previously appeared.
A more detailed and extensive
application of our criteria to supersymmetric extensions of the
standard model is in progress\cite{AC2}.

\section{Traditional Measures of Fine Tuning }
\def\theequation{2.\arabic{equation}}
\setcounter{equation}{0}
\indent

When parameters conspire by cancelling or adding in an unusually
precise fashion, we think of an atypical quantity that results
as fine tuned.
In such instances,
the quantity, for example $M_Z$, will exhibit a very strong dependence
on its arguments\cite{WtH}.
In supersymmetric extensions
of the standard model, the weak scale depends on the
soft supersymmetry breaking parameters and other couplings
through the renormalization group\cite{IR}.  In a seminal paper,
Barbieri and Giudice used  these features
to place upper bounds on superpartner masses, and they popularized
a prescription to quantify fine tuning that is now widely used.
They looked for sensitivity in the $Z$ mass to variations in the
values of supersymmetry breaking parameters and other couplings.
They measured the sensitivity on a general parameter $a$ by:
\begin{equation}
   c(M_Z^2;a) = | \frac{a}{M_{Z}^2}
   \frac{\partial M_{Z}^2}{\partial a} |\ .
\end{equation}
Note that rescaling the derivative by $a/M_{Z}^2$ removes the
dependence on the overall scale of $a$ and $M_Z$.
Barbieri and Giudice argued that, if supersymmetry is responsible for
stabilizing the weak scale, then $c(M_Z^2;a)$ must be less than some
upper limit $\Delta$, which they took to be $10$.
They used this criterion
to place upper limits on supersymmetry breaking parameters.
This program has been subsequently adopted by many
researchers.

The application of Eq.~(2.1) in obtaining upper bounds on superpartner
masses raises several questions.
Do we know the normalization of Eq.~(2.1) well enough to
say that natural solutions should exhibit $c(M_Z^2,a)$'s
below $10$ or any other particular value?
Should we expect that a simple application of this formula will
always give a reliable measure of fine tuning, and if not,
can we construct alternative definitions that provide better
measures of naturalness?
We can apply Eq.~(2.1) to a
famous hierarchy in order to shed some light on these questions.

The lightness of the proton in comparison to either the Planck scale
or the grand unified scale is beautifully explained by
the logarithmic running of the QCD coupling, $\alpha_3$.
At one loop, the scale dependence of the strong coupling constant can be
expressed as
\begin{equation}
   \alpha_{3}(\mu) = \frac{\\alpha_{3}(M_{Pl})}{1
   - \frac{b_3}{2\pi} \alpha_{3}(M_{Pl})
\ln(M_{Pl}/\mu)}.
\end{equation}
For simplicity we take $M_{prot} = \Lambda$, where
$ \alpha_{3}(\Lambda) = 1$.
A straight forward application of Eq.~(2.1) to the proton mass
yields
\begin{equation}
   c(M_{prot};g_s(M_{Pl})) = \left( \frac{4\pi}{b_3}\right)
\frac{1}{ \alpha_{3}(M_{Pl})} \gsim 100.
\end{equation}
The large value of $c(M_{prot},g_3(M_{Pl}))$
occurs because the proton mass is a very sensitive function of $g_3(M_{Pl})$.
The lightness of the proton is, of course, not the
result of a fine tuning.    The proton mass would have exhibited this
strong sensitivity no matter what its value was, so it makes
no sense to say that a value near $1$ GeV is fine tuned.
This example  illustrates our central point.  Equation (2.1)
is really a measure of sensitivity, and sensitivity does not
automatically translate into fine tuning.  For example,
the overestimate of fine tuning would have been even worse
had we used Eq.~(2.1) to study the naturalness of the technicolour
scale with respect to variations in the value of the technicolour
gauge coupling at the extended technicolour scale.
\footnote{In these examples there are no cancellations that we can
precisely adjust to create a large fine tuning.  However, even in
instances of real fine tuning, the largeness of $c(X;a)$ can be,
{\it in part}, due to global sensitivity.   As we will show in
section four, $c(M_{Z}^2 ;a)$ over estimates the amount of
fine tuning needed to maintain a light $Z$ mass in supersymmetric
extensions of the standard model.}

A reliable measure of fine tuning should
give a large value when a quantity
is fine tuned and at the same time reduce to
something close to
unity when it encounters typical sensitivity.
This suggests that we divide Eq.~(2.1) by some measure of average
sensitivity.  The resulting ratio will still be large for
solutions that are unusually sensitive, but in cases where
solutions have a ``typical'' sensitivity the
resulting ratio will be of order one.  So a more reliable measure
of fine tuning would be
\begin{equation}
\gamma(a) = c(X;a)/\bar{c},
\end{equation}
where $\bar{c}$ is some average value of $c(X;a)$.
For example,
\begin{equation}
\bar{c} = \frac{\int c(a) da}{\int da},
\end{equation}
or
\begin{equation}
1/\bar{c} = \frac{\int c^{-1}(a) da}{\int da} \ .
\end{equation}
If we apply this new criterion to the lightness of the proton,
we find that $\gamma$ is of order one.  It is a simple matter
to check that the ratio $\gamma$ gives a large
value in legitimate cases of fine tuning.  If we apply
Eq.~(2.4) to the weak scale hierarchy in a non-supersymmetric model,
we get a number of order $\Lambda/M_{weak}$, where $\Lambda$
is the scale of the cutoff.  As we show in the following
section, a ratio in a form of Eq.~(2.4) can be deduced from
very general considerations.

\section{Measuring Fine Tuning }
\def\theequation{3.\arabic{equation}}
\setcounter{equation}{0}
\indent

In this section we construct a family
of quantitative measures of fine tuning that encompass
Eq.~(2.4), the augmented
prescription we motivated in the previous section.  Our purpose
is twofold.  First, we wish to systematically
clarify what measures of fine tuning
best quantify our intuitive notion of naturalness and how
these measures should be normalized.  Second, we wish to make
explicit the inherent, discretionary assumptions present in any
standard that quantifies naturalness.
Any measure of fine tuning that quantifies naturalness
can be translated into an assumption about how likely
a given set of Lagrangian parameters is.
In the absence of a  theoretical reason compelling us to
choose a certain value, we can consider
some sensible distribution of the parameter to study what are the
natural predictions of the model.
The ``theoretical license''
at one's discretion when making this choice necessarily introduces
an element of arbitrariness to the construction.

Before we proceed to ``derive''
a quantitative measure of fine tuning some comments are
in order.  We are motivated to quantify naturalness
for tangible theoretical reasons.  A model
that explains a phenomenon has more predictive power than a model
that merely accommodates it.  In addition,
we understand why the proton can be naturally many orders of magnitude
lighter than the Planck scale but, the stability of a light scale
in a theory of fundamental scalars is mysterious.
We would like to understand how the weak scale remains light.
Of course, at the level of low energy effective theories,
dismissing ``unnatural'' theories
in the quest for a ``natural'' explanation of weak scale stability
could be misguided.  We certainly cannot prove that
an explanation of the light weak scale was not butchered by
the process in which we constructed our effective theory.
For example, one loop corrections
to the cosmological constant from an effective theory with soft
supersymmetry breaking generate
contributions that are many orders of magnitude greater than the
experimental limit.  Yet we often entertain the idea that the solution
to this problem
is not associated with our choice of a low energy Lagrangian.
While we cannot elevate the prejudice of searching for natural
theories above the level of an axiom,
we can hope that its application will lead us to a more complete
model that explains the stability of the weak scale.
Such models will have testable predictions.

  In light of this, we proceed to deduce a measure of fine tuning
from general principles.  Provided we parameterize our
assumptions about the likely distribution for Lagrangian parameters,
we should be able to derive a quantitative measure of naturalness.
Assume the probability that a Lagrangian parameter lies between
$a$ and $a + da$ is
\begin{equation}
dP(a) = \frac{f(a) da}{\int f(a) da}.
\end{equation}
Consider a  set of these
Lagrangian parameters $a_i$ specified at a renormalization scale
that is the high energy boundary of our effective theory
({\it e.g.}, $\mu = M_{GUT}$).  A measurable parameter $X$
({\it e.g.}, $M_Z^2$) will depend on the $a_i$ through the
renormalization
group equations and possibly on a set of minimization conditions.
We can recast Eq.~(3.1) as a probability per
unit $X$.
Given a probability density $f(a)$, the probability
density per unit $X$ is
\begin{equation}
dP = \rho(X) \,dX,
\end{equation}
where
\begin{equation}
\rho(X) \simeq \frac{1}{X\,c(X;a)}
\frac{a f(a)}{\int f(a)da}.
\end{equation}

In studies of naturalness, we may ask:  If the fundamental Lagrangian
parameters at our high energy boundary condition are
distributed like $f(a)$, how likely is  a low energy observable,
$X(a)$, to be contained in an interval $u(X)\,dX$ about $X$?
A quantity $X$ is relatively unlikely to be in an interval
proportional to $u(X) dX$  if
\begin{equation}
 \frac{<u\rho>}{u(X)\rho(X)} >> 1,
\end{equation}
where $<u\rho> = \int da\, u(X)\rho(X)/ \int da$.

If we are interested in studying the naturalness of a hierarchy
like $M_{weak}/M_{GUT}$, $M_{prot}/M_{Planck}$, or
$M_{Z}^2/M_{SUSY}^2$, the interval that corresponds to
the conventional sense of naturalness is $u(X) = X$.
\footnote{Consider the hierarchy problem in an effective
theory with a fundamental scalar defined below some scale
$\Lambda_1$: $m_{S}^2 = g^2 \Lambda_{1}^2 - \Lambda_{2}^2$.
The scalar mass
can only remain light in comparison to the cutoff scale $\Lambda_1$
if we cancel the quadratic divergence against the bare term
$\Lambda_{2}^2$.  Note that
the cancellation we need to place the scalar mass in a 1 GeV window
at $10^{16}$ GeV must be made with the same precision
as the cancellation we need to place the scalar mass in a 1 GeV
window at a 100 GeV. A small value of the scalar mass is
unnatural in the  sense that a small change in $g$ leads to a large
{\it fractional} change in $m_{S}^2$ so that it is relatively
unlikely to be found in an interval $\propto m_{S}^2 dm_{S}^2$
around $m_{S}^2$.
}

If we define our measure of naturalness as
\begin{equation}
\gamma = <X\rho>/X\rho ,
\end{equation}
fine tuning corresponds to $\gamma >> 1$.
The definition of $\gamma$ in Eq.~(3.5) necessarily
implies that $\gamma$ is linearly proportional to $c$.
For any realization of $\gamma$, we define
an average value of $c(X;a)$ by
\begin{equation}
\gamma = c/\bar{c} .
\end{equation}
This definition of $\bar{c}$ corresponds to
\begin{equation}
   \bar{c}^{-1} = \frac{\int da \, a f(a) c(X;a)^{-1}}
{a f(a)\int da}.
\end{equation}
The similarity between this definition of $\bar{c}$
and the heuristic average posed in section two is
apparent.

In order to make practical use of the prescription
contained in Eqs.~(3.4)-(3.7), we need to specify three
things.  First, our choice of $f(a)$ reflects our theoretical
prejudice about what constitutes a natural value of the Lagrangian
parameter $a$.  We will return to this point in section four.
The two remaining choices are determined by the
questions we wish to ask.  Our choice of $u(X)$ is determined by
the quantity whose naturalness we wish to study.  The conventional
notion of naturalness for hierarchy problems suggests
$u(X) = X$. Finally,
our choice for the range of integration for $a$
is related to the
broadness of the question we wish to ask.
This point will be elaborated upon in section four.

Before analyzing the naturalness of radiative symmetry breaking
in the supersymmetric standard model we
specialize Eqs.~(3.6) and (3.7) to two examples.\\
$\bullet$ {\bf Example I} \\
Let's return to the hierarchy between the proton mass and the
Planck scale discussed in section two.  We will calculate
$\gamma$ for two different choices for $f(a)$.
Integrating over $g_s(M_{Pl})$ in the range $g_{-} < g < g_{+}$
we find
\begin{equation}
   \gamma_1 = \left(\frac{g_{+}+g_{-}}{4g}\right)
\frac{g_{+}^2 + g_{-}^2}{g^2},
\end{equation}
for $f(g) = 1$ and
\begin{equation}
\gamma_2 = \frac{1}{3}
\left(\frac{g_{+}^2 + g_{+}g_{-} + g_{-}^2}{g^2}\right),
\end{equation}
for $f(g) = 1/g$.  In each case we see that, if the
strong coupling constant at the Planck scale is of order
one, our measure indicates that a $1$ GeV proton mass
arises naturally.  We have thus eliminated the problematic
overestimate of fine tuning contained in the traditional
prescription.  In the following example we show that
the new
prescriptions still registers appropriately
large values in real
instances of fine tuning.

$\bullet$ {\bf Example II}\\
Consider the gauge hierarchy problem in a non-supersymmetric
theory with fundamental scalars.  In this case, the one-loop
correction to the  scalar mass
will be of the form
\begin{equation}
   m_{S}^2(g) = g^2 \Lambda_{1}^2 - \Lambda_{2}^2,
\end{equation}
where $\Lambda_1 $ is the ultraviolet cutoff of our effective
theory, and $\Lambda_2$ is a bare term chosen to keep the scalar
mass light.
If we calculate the sensitivity of the Higgs mass with respect
to the coupling $g$, we find
\begin{equation}
   c(M_{S}^2;g) = 2\frac{g^2 \Lambda_{1}^2}{m_{S}^2(g)}.
\end{equation}
Integrating over $g$ in the range $g_{-} < g < g_{+}$, we find
\begin{equation}
   \bar{c}_1^{-1} = \frac{1}{g(g_{+}-g_{-})}
\left(\frac{1}{2}\right)
   \left[\frac{1}{2}\left( g_{+}^2 - g_{-}^2\right) -
   \frac{\Lambda_{2}^2}{\Lambda_{1}^2}
   \ln \left(\frac{g_{+}}{g_{-}}\right)\right]
\end{equation}
for $f(g) = 1$ and
\begin{equation}
\bar{c}_2^{-1} =\frac{1}{2}\left[1 - \frac{1}{g_{+}g_{-}}
 \left( \frac{\Lambda_{2}^2}{\Lambda_{1}^2} \right) \right],
\end{equation}
for $f(g) = 1/g$.  In each case $\bar{c}$ is of order one, while
$c(m_S^2;g)$ is of order $\Lambda^2 / m_{S}^2$. This gives
$\gamma \simeq \Lambda^2/ m_{S}^2$, which correctly
reproduces the fine tuning needed to maintain light scalar
masses.  From these examples, we again see that
the need to renormalize $c(X;a)$ by $\bar{c}$ is important.  When
$X$ depends very sensitively on $a$, $c(X;a)$ will be large
even if there is no fine tuning.  A largely exaggerated value
for the traditional fine-tuning measure, which can occur in the
absence of fine tuning, can be removed by rescaling by $\bar{c}$.

\section{Naturalness and the MSSM}
\def\theequation{4.\arabic{equation}}
\setcounter{equation}{0}
\indent

There are two issues concerning naturalness that should be addressed
for radiative electroweak symmetry breaking (EWSB) in supersymmetric
extensions of the standard model.
The first concerns the natural value of the electroweak scale if
electroweak symmetry breaks.  The
second concerns the naturalness of the EWSB process itself.
We will make no attempt to tackle the second problem in this paper,
since this would require either knowledge of, or additional assumptions
about, a more complete theory.

As already noted, supersymmetry must be broken to reconcile the MSSM
with the lack of experimental evidence for superparticles.  Since no
adequate model of spontaneously broken global SUSY exists, supersymmetry
is customarily broken through the introduction of explicit soft terms
that do not reintroduce quadratic divergences into the theory.  Low
energy supergravity provides the motivation for the introduction of
these soft breaking terms.  The most general form of the soft SUSY
breaking potential, including gaugino mass terms, is
\begin{eqnarray}
   V_{\rm soft} &=&  m_{\Phi_u}^2 | \Phi_u |^2
                   + m_{\Phi_d}^2 | \Phi_d |^2
                   + B \mu ( \Phi_u \Phi_d + h.c. )
                     \nonumber \\
   & &\mbox{}      + m_{\tilde Q}^2 | {\tilde Q}_i |^2
                   + m_{\tilde L}^2 | {\tilde L}_i |^2
                   + m_{\tilde{\overline u}}^2 | {\tilde{\overline u}} |^2
                   + m_{\tilde{\overline d}}^2 | {\tilde{\overline d}} |^2
                   + m_{\tilde{\overline e}}^2 | {\tilde{\overline e}} |^2
                     \nonumber \\
   & &\mbox{}      + A_u Y_u {\tilde{\overline u}} \Phi_u {\tilde Q}
                   + A_d Y_d {\tilde{\overline d}} \Phi_d {\tilde Q}
                   + A_e Y_e {\tilde{\overline e}} \Phi_d {\tilde L}
                   + {1\over2} M_l \lambda_l \lambda_l + h.c. \ .
                     \label{vsoft}
\end{eqnarray}
A generic feature of these SUGRA inspired models is universality in the
soft terms.  Therefore, one customarily assumes the following boundary
conditions for the masses and trilinears at the gauge coupling unification
scale
\begin{equation}
   m_i = m_0 \ , \;\;\;
   A_u = A_d = A_e = A_0 \ .
\label{bcm0a}
\end{equation}
Some universality is important in avoiding unwanted flavor changing
neutral current effects.
Given the unification of gauge couplings, it is natural to take the gaugino
masses equal as well
\begin{equation}
   M_1 = M_2 = M_3 = m_{1/2} \ .
\label{mg}
\end{equation}
There are therefore five soft breaking parameters, $m_0$, $A_0$, $m_{1/2}$,
$B_0$, and $\mu_0$ in the simplest version of the MSSM.
For simplicity and definiteness, we will concentrate
on this restricted version of the minimal supersymmetric standard model
in this paper, however, our naturalness criteria apply equally well to
other scenarios.

In the MSSM, electroweak symmetry breaking proceeds through radiative
effects\cite{IR,IKKT,AGPW,EHNT}.  The 1-loop effective Higgs potential may
be expressed as follows
\begin{equation}
   V_{1-{\rm loop}}(Q) = V_0(Q) + \Delta V_1(Q) \ ,
\label{vhiggs}
\end{equation}
where $V_0$ is the tree level potential, and $\Delta V_1$ represents
the 1-loop correction.~\footnote{The effect of including
one-loop corrections to the effective potential on the numerical
value of the Barbieri-Giudice parameter $c(M_{Z}^{2},y_t)$ was studied in
Ref. [9].}
Using the renormalization group, the parameters are evolved to low energies
where the potential attains validity.  This renormalization group improvement
uncovers electroweak symmetry breaking.  The exact low energy scale at which
to minimize is unimportant as long as the 1-loop effective potential is used
and the scale is in the expected electroweak range.  If we arbitrarily take
the minimization scale to be $M_Z$, then the two minimization conditions may
be expressed as follows
\begin{eqnarray}
   \mu^2(M_Z) &=& {{\overline m}_{\Phi_d}^2 -
   {\overline m}_{\Phi_u}^2 \tan^2\beta \over \tan^2\beta - 1 }
   - {1\over2} M_Z^2 \ , \label{min1} \\
   B(M_Z) &=& { ({\overline m}_{\Phi_u}^2 + {\overline m}_{\Phi_d}^2
              + 2\mu^2)\sin2\beta \over
              2 \mu(M_Z) } \ , \label{min2}
\end{eqnarray}
where ${\overline m}_{\Phi_{u,d}}^2 = m_{\Phi_{u,d}}^2 +\partial\Delta
V_1/\partial v_{u,d}^2$ and $\tan\beta$ is the ratio of the vacuum
expectation values of the Higgs fields, $v_u/v_d$.
Demanding correct electroweak symmetry breaking
puts constraints on the parameters of the MSSM.  For example,
the top quark Yukawa coupling
is one parameter that has to be large enough in order to achieve the desired
radiative breaking.  Rewriting Eq.~(4.5) yields an equation for
$M_Z$ as a function of the parameters of the MSSM
\begin{equation}
   {1\over2} M_Z^2 = {{\overline m}_{\Phi_d}^2 -
   {\overline m}_{\Phi_u}^2 \tan^2\beta \over \tan^2\beta - 1 }
   -  \mu^2 \ .
\label{mz2}
\end{equation}

In the MSSM, the problem of fine tuning has been commonly treated using the
prescription of Ref. [1], although the original bound of $\Delta=10$
has often been increased to as high as $\Delta=100$.  However, as
already discussed, it is difficult to ascertain what constitutes a
reasonable bound in the absence of some comparative norm (normalization).
A glaring example of this can be found in $c(M_Z^2;a=g_3)$.  When
applying the criterion of Ref. [1], one typically takes the $a$-parameter
to be a soft breaking mass, such as $m_0$, $m_{1/2}$, $\mu_0$, etc., or
the top Yukawa.  However, the strong coupling is also a parameter of the
theory, and one can consider $c(M_Z^2;g_3)$.  We find that over all
the parameter space of the MSSM that we have so far explored,
$c(M_Z^2;a)$ is the largest for $a=g_3$.  Since all the parameters are
ostensibly on equal footing, imposing $c(M_Z^2;g_3)<\Delta=10-100$ may
be overly restrictive.

We now apply the realization of $\gamma$ given in
Eqs.~(3.5)-(3.7) to the MSSM.  To use this prescription
we must specify the range of the parameter $a$.
We could simply choose this range by {\it fiat} ({\it e.g.},
$ 0 <m_{1/2} < 10$ TeV), but this seems rather {\it ad hoc}.
Instead we prefer that the choice of range be dictated by
electroweak symmetry breaking.   Other choices are possible.
We will ask how natural the
value $M_Z=91.2$ GeV is, given that the gauge symmetry breaking occurs
correctly.   For this choice,
the range of $a$ should correspond to values for which
$SU(3)\times SU(2) \times U(1)$ is broken to $SU(3) \times U(1)_{em}$.
There are then finite limits to the range of $a$ that come from two conditions
on the value of $M_Z$.  The minimum value of $M_Z$ cannot be less than $0$,
and its maximum value cannot exceed some upper bound, often set by the
requirement that sneutrino squared masses be positive.

We display $\gamma(a)$'s computed for two different choices of $f(a)$.
$\gamma_1$ corresponding to the choice $f(a)=1$, and $\gamma_2$ corresponding
to $f(a)=1/a$.
If we adopt 't Hooft's notion of naturalness that Lagrangian parameters
should not be small unless setting them to zero increases a symmetry,
and we believe that the magnitude of supersymmetry breaking terms should be
universal, we should choose $f(a)=1$.
However, we also consider $f(a) = 1/a$ to study the sensitivity of our
criteria to the choice of $f(a)$ and to allow for
non-universality in the magnitude of soft supersymmetry breaking terms
(see for example Ref. [8]).
Figures 1-3 show that, in the MSSM, the $\gamma$'s are very
insensitive to which choice of $f(a)$ is made.

In Figs. 1a and 1b, we plot $\gamma(m_{1/2})$ vs. $m_{1/2}$
for two choices of the soft supersymmetry breaking parameters
$A_0$, $B_0$, $m_0$, and $\mu_0$.
On this scale, $\gamma_1$ and $\gamma_2$ are virtually
indistinguishable so we only show $\gamma_1$.
On the same plots we show, for comparison, the traditional prescription
$c(m_{1/2})$ as well.  The range of $m_{1/2}$ corresponds to the values
of the common gaugino mass for which $SU(3)\times SU(2) \times U(1)$ is
broken to $SU(3) \times U(1)_{em}$.   Note that the asymptotic,
``natural'' value of $c(m_{1/2})$ for large $m_{1/2}$ is order ten and
not order one.  This is another demonstration why it is necessary to
rescale the $c$'s to achieve a sensible measure of fine tuning.

Figures 2a and 2b show the effect of increasing the overall scale of
soft symmetry breaking on fine tuning.  In Fig. 2a the fine-tuning parameter
$\gamma(g_3)$ is plotted as a function of $g_3(M_U)$, where $M_U$ is the
unification scale.  We include three choices of $A_0$, $B_0$, $m_0$, and
$\mu_0$ with different overall scales of soft symmetry breaking.  The square,
circle, and diamond in each figure correspond to the point with the correct
value of the Z-boson mass for the cases (i) $A_0=m_0=m_{1/2}=400$ GeV,
$B_0=523$ GeV, $\mu_0=1125$ GeV, (ii) $A_0=m_0=m_{1/2}=200$ GeV,
$B_0=275$ GeV, $\mu_0=585$ GeV, and (iii)
$A_0=m_0=m_{1/2}=50$ GeV, $B_0=90$ GeV, $\mu_0=154$ GeV, respectively.
The light case has a chargino with a mass less than $M_Z/2$ and
therefore is excluded experimentally.
Figure 2b is similar to 2a but displays $\gamma(y_t)$ vs. $y_t(M_U)$.

Fig.~3 displays how much fine tuning the MSSM currently requires
in light of some general experimental constraints.  We consider a region
of our input parameter space defined by the ranges $|A_0|\leq 400$ GeV,
$m_0\leq 400$ GeV, and $|m_{1/2}|\leq 400$ GeV.  For values of the soft
supersymmetry breaking parameters consistent with a neutral lightest
SUSY particle (LSP), with the current LEP measurement of the $Z$ width,
with a Higgs mass heavier that $60$ GeV, and with chargino masses heavier
than $M_{Z}/2$, 
we plot ${\tilde c}={\rm max}\{c(m_{1/2}),c(m_0),c(y_t),c(g_3)\}$,
${\tilde\gamma}_1={\rm max}\{\gamma_1(m_{1/2}),\gamma_1(m_0),\gamma_1(y_t),
\gamma_1(g_3)\}$, and ${\tilde\gamma}_2={\rm max}\{\gamma_2(m_{1/2}),
\gamma_2(m_0),\gamma_2(y_t),\gamma_2(g_3)\}$ {\it vs.} $\tan\beta(M_Z)$.
In the figure, we display curves representing the lower envelopes of the
resulting regions.   Notice that the original Barbieri and Giudice
bound of $c(a) < 10$ has already been exceeded,
while the new criteria show that weak scale stability can still arise
naturally.

Finally, in Table 1 we display  the BG sensitivity parameters $c(a)$
and the fine tuning parameters $\gamma(a)$
for various $a$ in a representative case with $A_0=m_0=m_{1/2}=200$ GeV,
$B_0=275$ GeV, and $\mu_0=585$ GeV.  Note that the relative
normalization of
the sensitivity parameters, $\bar{c}(a)$, can be quite different.
This means that we can not adopt a universal measure of fine tuning
by appealing only to the $c(a)$'s ({\it e.g.}, $c<100$).  A relative
normalization for each $c(a)$ must be computed in the manner
described in section three.

\begin{center}
\begin{large}
\indent{\bf Table 1}
\vskip 20pt
\begin{tabular}{|c|c|c|c|c|c|}
\hline
$a$ & $c(a)$ & $\bar{c}_1$ & $\bar{c}_2$ & $\gamma_1$ & $\gamma_2$ \cr
\hline
\hline
$m_{1/2}$ & $50.8$ & $9.29$ & $10.3$ & $5.47$ & $4.92$\cr
\hline
$m_{0}$   & $ 21.8 $ & $3.21$ & $4.66$ & $6.79$ & $4.68$\cr
\hline
$g_3$   & $ 209. $ & $42.3$ & $43.2$ & $4.94$ & $4.84$\cr
\hline
$y_t$   & $ 32.5 $ & $4.92$ & $5.77$ & $6.61$ & $5.63$\cr
\hline
\end{tabular}
\end{large}
\end{center}

\def\theequation{5.\arabic{equation}}
\setcounter{equation}{0}
\section{Conclusions}
\indent
   Naturalness criteria are frequently used to place upper
bounds on superpartner masses in supersymmetric
extensions standard model. We have analyzed the prescription
popularly used to measure fine tuning.  This prescription is
an operational implementation of Susskind's statement of
Wilson's sense of naturalness, ``Observable properties of a system
should be stable against minute variations of the fundamental
parameters.'' Because this prescription is only
a measure of sensitivity, we found that it is not a reliable
measure of naturalness.  We then constructed a family of
prescriptions which measure fine tuning more reliably.
Our measure is an operational implementation of a modified
version of Wilson's naturalness criterion: Observable properties
of a system should not be unusually unstable against minute variations
of the fundamental parameters. Our derivation
determines the normalization of naturalness measures and
makes clear to what extent theoretical prejudice influences
these measures.  The new prescriptions we construct allow upper
bounds on
superpartner masses to be placed with greater confidence.
By applying our prescriptions to the minimal supersymmetric
standard model, we find that the theory provides a
much more natural explanation of weak scale stability than
previous methods indicate.  More importantly, we find that the
scale of supersymmetry breaking can be significantly higher
than previous naturalness criteria indicate.

\section*{Acknowledgments}
GA would like to thank the Institute for Theoretical
Physics in Santa Barbara and the
Aspen Center for Physics for their hospitality.

\newpage
\begin{description}

\item[\it Figure 1a:] The fine-tuning parameters $c(m_{1/2})$ (solid) and
$\gamma(m_{1/2})$ (dashed) plotted as a function of $m_{1/2}$ for
$A_0=m_0=200$ GeV,
$B_0=275$ GeV and $\mu_0=585$ GeV.  The circles indicate the point
with the correct value of $M_Z$ for this choice of $A_0$, $B_0$, $m_0$,
and $\mu_0$.

\item[\it Figure 1b:] Same as Figure 1a with $A_0=m_0=100$ GeV,
$B_0=143$ GeV and $\mu_0=305$ GeV.

\item[\it Figure 2a:] The fine-tuning parameter $\gamma(g_3)$
plotted as a function of $g_3(M_U)$ for three cases with increasing
scale of supersymmetry.  The circle, square, and diamond indicate the points
with the correct value of $M_Z$ for the three cases.

\item[\it Figure 2b:] Same as Figure 2a but displays $\gamma(y_t)$
as a function of $y_t(M_U)$ for the same three cases.

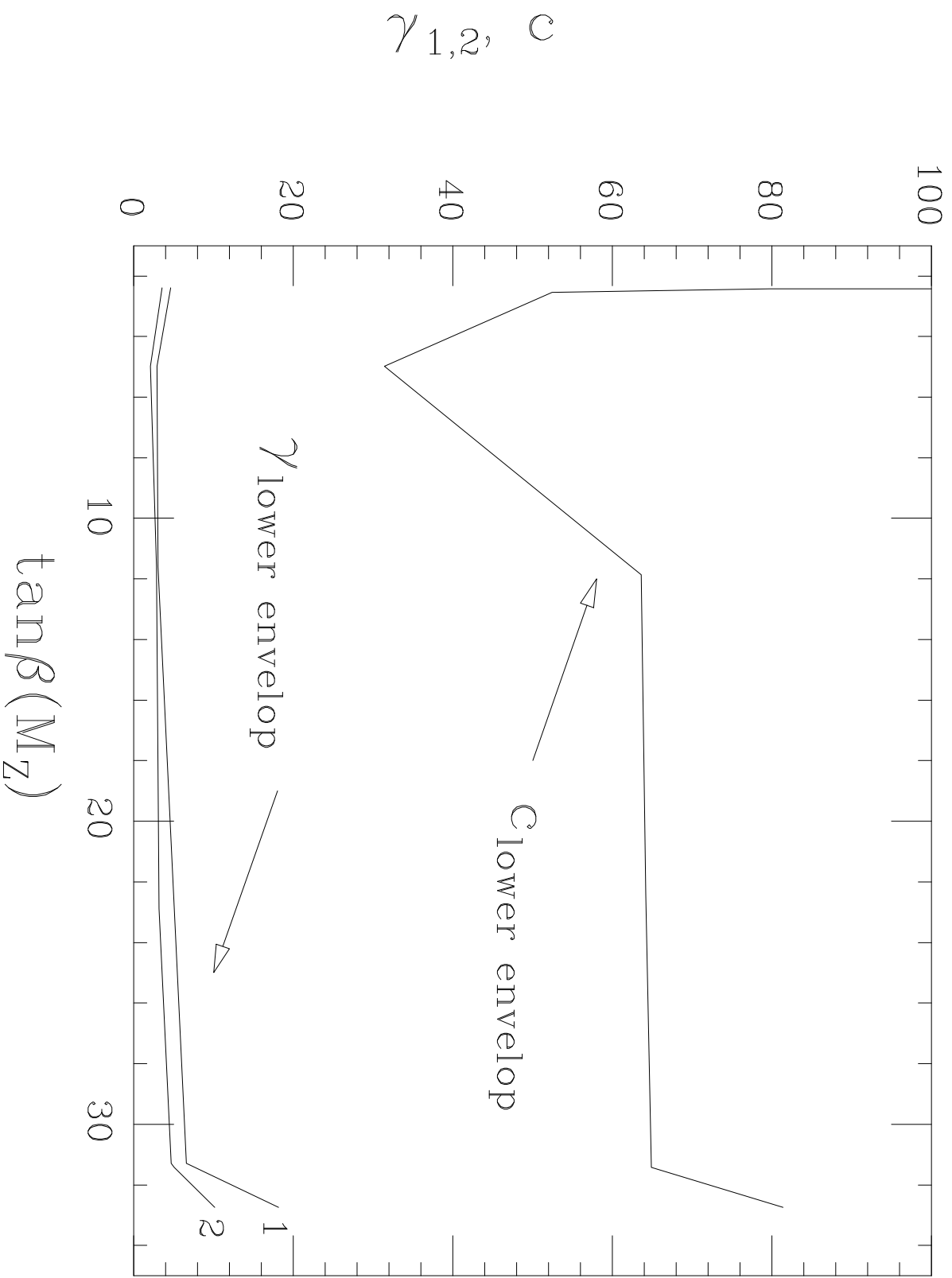
\item[\it Figure 3:] Curves representing lower envelope of regions defined
by plot of ${\rm max}\{c(m_{1/2}),c(m_0),c(y_t),c(g_3)\}$ and
${\rm max}\{\gamma_{1,2}(m_{1/2}),\gamma_{1,2}(m_0),\gamma_{1,2}(y_t),
\gamma_{1,2}(g_3)\}$ vs. $\tan\beta(M_Z)$.

\end{description}

\begin{thebibliography}{99}

\bibitem{BG} R. Barbieri and G. F. Giudice, Nucl. Phys,
{\bf B306}, 63 (1988);
J. Ellis, K. Enqvist, D.V. Nanopoulos, and F. Zwirner,
Mod. Phys. Lett. {\bf A1}, 57 (1986).
\bibitem{WtH} K. Wilson, as quoted by L. Susskind, Phys. Rev.
{\bf D20}, 2619 (1979); G.'t Hooft, in {\it Recent developments
in gauge theories}, ed by G. 't Hooft et al. (Plenum Press, New
York, 1980)p. 135
\bibitem{IR}L. Ib\'a\~nez and G.G. Ross, Phys. Lett {\bf 110B}, 215 (1982)
\bibitem{IKKT}K.~Inoue, A.~Kakuto, H.~Komatsu, and S.~Takeshita,
Prog. Theor. Phys. {\bf 68}, 927 (1982).
\bibitem{AGPW} L.~Alvarez-Gaum\'e, M.~Claudson, and M.~B.~Wise,
Nuc. Phys. {\bf B207}, 96 (1982);
L.~Alvarez-Gaum\'e, J.~Polchinski, and M.~B.~Wise,
Nuc. Phys. {\bf B221}, 495 (1983).
\bibitem{EHNT}J.~Ellis, J.~S.~Hagelin, D.~V.~Nanopoulos, and K.~Tamvakis,
Phys. Lett. {\bf 125B}, 275 (1983).
\bibitem{AC2} G. W. Anderson and , D. J. Casta\~no, Phys. Rev. {\bf D52},
1693-1700 (1995) hep-ph/9412322 ; MIT-CTP-2464, hep-ph/9509212.
\bibitem{Louis}V.S. Kaplunovsky and J. Louis,
Phys. Lett. {\bf B306}, 269 (1993).
\bibitem{CC}B. de Carlos and J.A. Casas, Phys. Lett {\bf B309}, 320
(1993).
\end{thebibliography}
\end{document}